\documentstyle[preprint,aps,epsfig]{revtex}

\begin{document}
 
\draft

\bibliographystyle{prsty}

\title
{\bf FORM FACTORS OF THE NUCLEON \\
IN THE SU(3) CHIRAL QUARK-SOLITON MODEL 
}
\author{Hyun-Chul Kim\footnote{Invited talk given at 
Workshop on Electron-Nucleus Scattering,
 Elba International Physics Center, 1-5 July 1996 }}
\address{Institut f\"ur Theoretische Physik II \\
Ruhr-Universit\"at Bochum \\
D-44780 Bochum  Germany}

\date{August, 1996}
\preprint{RUB-TPII-13/96} 
\maketitle

\begin{abstract}  
The recent investigation on various form factors of the nucleon is
reviewed in the framework of the SU(3) chiral quark-soliton model.  
The results for the electromagnetic and scalar form factors are
in remarkable agreement with experimental and empirical data.
The strange vector form factors are also discussed with the effect of the
kaon cloud being considered.  
In addition to the form factors, the recent calculation
of the tensor charges is presented.
\end{abstract}
 
\pacs{Pacs number(s): 11.15.Pg, 12.40.-y, 13.40.Gp, 14.20.Dh\\
Keywords: Electromagnetic form factors, scalar form factors, 
strange vector form factors, tensor charges, chiral quark-soliton model}
 

\narrowtext

\section{Introduction}
Though Quantum Chromodynamics (QCD) is believed to be the underlying 
theory of the strong interactions, low energy phenomena such as static 
properties of hadrons defy solutions based on QCD because of formidable
mathematical complexities.  The pertinacity of QCD in the low energy
regime have led to a great deal of efforts to construct an effective
theory for the strong interactions.  In pursuit of this aim, the chiral
quark-soliton model ($\chi$QSM), 
also known as the semibosonized Nambu-Jona-Lasinio
model, emerged as a simple and successful effective theory to describe
the low energy phenomena without loss of important properties of
QCD such as chiral symmetry and its spontaneous breaking.  

Originally, the idea of finding the soliton in a model with quarks
coupled to pions was realized by Kahana, Ripka and Soni~\cite{KaRipSo} 
and Birse and Banerjee~\cite{BirBa}.  
 The bound states of the valence quarks were well explored
in the model while it suffered from the vacuum instability.  

Having studied the QCD instanton vacuum in the low-momenta limit, 
Diakonov and Petrov obtained an effective Lagrangian resembled  
the Nambu-Jona-Lasinio type model~\cite{njl} with $2N_f$-quark vertices.
They showed that the resulting bosonized low-momenta 
theory is equivalent to the $\chi$QSM free from the vacuum instability.  
In fact, the effective model by Diakonov and Petrov has several important
virtues: The mechanism of the spontaneous breaking of chiral symmetry
is well explained in a natural way.  It provides also a renormalization
scale by the inverse of the average size of the instanton $1/\rho\sim 600$
MeV.  In addition, the $\chi$QSM throws a bridge between the 
nonrelativistic constituent quark model and the Skyrme model.

The baryon in this model is regarded as $N_c$ valence quarks coupled to
the polarized Dirac sea bound by a nontrivial chiral field configuration
in the Hartree approximation~\cite{dpp,RW,mgg,WY}.  The identification
of the soliton as the baryon is acquired by the semiclassical collective
quantization~\cite{dpp,anw} which is performed by integrating over 
zero-mode fluctuations of the pion field around the saddle point.
The model enables us to describe quantitatively a great deal of
static properties of the nucleon such as baryon octet-decuplet mass 
splittings~\cite{Blotzetal}, magnetic moments~\cite{KimPolBloGoe},
axial constants~\cite{WW,Chretal1}, electromagnetic form 
factors~\cite{Chretal2,Kimetal} and so on (see for example a recent
review~\cite{Review}).    

\section{General formalism}
Let me first sketch the $\chi$QSM.  Because one can find the 
detailed formalism elsewhere, I want to present the general
idea of the model briefly.  

The $\chi$QSM in SU(3) is characterized by a low-energy partition
function in Euclidean space given by the functional integral
over pseudoscalar meson and quark fields:
\begin{equation}
{\cal Z}\;=\; \int {\cal D} \Psi {\cal D}\Psi^\dagger {\cal D}\pi
\exp{\left(-\int d^4x \Psi^\dagger iD \Psi\right)},
\label{Eq:Z}
\end{equation}
where $iD$ stands for the Dirac differential operator
\begin{equation}
iD\;=\;\beta \left(-i\rlap{/}\partial  + \hat{m} + MU^{\gamma_5}\right)
\end{equation}
with the pseudoscalar chiral field
\begin{equation}
U^{\gamma_5} \;=\; \exp{i\pi^a\lambda^a \gamma_5}.
\end{equation}
The $\hat{m}$ denotes the matrix element of the current quark mass given
by 
\begin{equation}
\hat{m}\;=\;\mbox{diag}(m_u,m_d,m_s) \;=\; m_0 {\bf 1} + m_3 \lambda_3
+ m_8 \lambda_8 .
\end{equation}
$\lambda^a$ represents the usual Gell-Mann matrices normalized as
$\mbox{tr} (\lambda^a \lambda^b)=2\delta^{ab}$.  
$M$ designates the dynamical quark mass arising from the spontaneous
breaking of chiral symmetry, which is in general momentum-dependent
\cite{dp}.  For convenience, we shall regard $M$ as a constant and
introduce the ultraviolet cutoff via the proper time regularization.
The $m_0$, $m_3$ and $m_8$ are respectively defined by
\begin{equation}
m_0\;=\;\frac{m_u+m_d+m_s}{3},\;\;\;m_3\;=\;\frac{m_u-m_d}{2}\;\;\;
m_8\;=\;\frac{m_u+m_d-2m_s}{2\sqrt{3}}
\end{equation}
We assume isospin symmetry, {\em i.e.} $m_3$ is taken to be zero.
The operator $iD$ is expressed in Euclidean space in terms of the
Euclidean time derivative $\partial_\tau$ and the Dirac one-particle
Hamiltonian $H(U^{\gamma_5})$ 
\begin{equation}
iD\;=\; \partial_\tau + H(U^{\gamma_5}) + \beta \hat{m} 
- \beta \bar{m}{\bf 1}
\end{equation}
with
\begin{equation}
H(U^{\gamma_5})\;=\; -i {\bf \alpha}\cdot \nabla + \beta MU^{\gamma_5}
+ \beta\bar{m} {\bf 1}.
\end{equation}
$\beta$ and ${\bf \alpha}$ are the well-known Dirac Hermitian matrices.
The $\bar{m}$ is defined by $(m_u+m_d)/2=m_u=m_d$, 
which is introduced to avoid certain divergences in some observables
such as the isovector electric charge radius 
in the chiral limit ($m_\pi\rightarrow 0$).

Note that the effective chiral action given by Eq.(\ref{Eq:Z}) 
contains the Wess--Zumino and four-derivative
Gasser--Leutwyler terms with correct coefficients 
in the gradient expansion.  
Therefore, at least the first four terms in 
the gradient expansion of Eq.(\ref{Eq:Z}) are correctly reproduced 
and chiral symmetry arguments do not leave much room for 
further modifications.  

In order to calculate an observable in the $\chi$QSM,
we consider a correlation function: 
\begin{equation}
\langle 0 | J_N ({\bf x}, \frac{T}{2})\bar{\Psi}\hat{\Gamma} \hat{O}\Psi 
J^\dagger_{N} ({\bf y}, -\frac{T}{2}) |0\rangle
\label{Eq:corr}
\end{equation}
at large Euclidean time $T$.  
$\hat{\Gamma}$ stands for the corresponding
spin operator, while $\hat{O}$ is the corresponding flavor operator.
The $J_N(J^\dagger_{N})$ denotes the nucleon current consisting of
$N_c$ quark fields.
The corresponding matrix element can be represented
by the Euclidean functional integral:
\begin{eqnarray}
\langle N, p' | \hat{\Gamma}\hat{O} | N, p \rangle & = & 
\frac{1}{\cal Z} \lim_{T \rightarrow \infty} \exp{(ip_4 \frac{T}{2}
- ip'_{4} \frac{T}{2})} \nonumber \\
& \times & \int d^3 x d^3 y 
\exp{(-i \vec{p'} \cdot \vec{y} + i \vec{p} \cdot \vec{x})} 
\int {\cal D}U^{\gamma_5} \int {\cal D} \Psi \int {\cal D}\Psi^\dagger 
\nonumber \\
& \times & \; J_{N}(\vec{y},T/2)\Psi^\dagger(0) 
\beta \hat{\Gamma} \hat{O} \Psi (0) J^{\dagger}_{N} (\vec{x}, -T/2) 
\nonumber \\ & \times &
\exp{\left[ - \int d^4 z \Psi^\dagger i D \Psi \right ]}.
\label{Eq:ev} 
\end{eqnarray}
With the quark fields being integrated out, Eq.(\ref{Eq:ev})
can be divided into two separate contributions:
\begin{equation}
\langle N, p'| \hat{\Gamma}\hat{O} | N, p \rangle 
\;=\;\langle N, p'| \hat{\Gamma}\hat{O} |N, p \rangle_{val}
\;+\;\langle N, p'| \hat{\Gamma}\hat{O} |N, p \rangle_{sea}.
\end{equation}
Schematically, the valence and sea contributions are shown in Fig.1.

The integral over chiral bosonic fields $U^{\gamma_5}$ can be carried out 
by the saddle point method in the large $N_c$ limit
with the following Ansatz: 
\begin{equation}
U\;=\; \left ( \begin{array}{cc}
U_0 & 0 \\ 0 & 1 \end{array} \right ),
\label{Eq:imbed}
\end{equation}
where $U_0$ is the SU(2) chiral background field 
\begin{equation}
U_0\;=\;\exp{[\vec{n}\cdot\vec{\tau}P(r)]} .
\label{Eq:profile}
\end{equation}
$P(r)$ denotes the profile function satisfying the boundary condition
$P(0) = \pi$ and $P(\infty)=0$.

Since the angular velocity $\Omega_E$ ($\sim 1/N_c$) 
and the strange quark mass $m_s$ are regarded as 
small parameters in our model, we expand the propagator
$1/i\tilde{D}$ with respect to the $\Omega_E$ and $m_s$ up to
the first order:
\begin{equation}
\frac{1}{i\tilde D}\;\approx \;
\frac{1}{\partial_\tau + H}
\;+\; \frac{1}{\partial_\tau + H} (-i \Omega_E) 
\frac{1}{\partial_\tau + H}   
\; + \; \frac{1}{\partial_\tau + H} (-\beta A^{\dagger} \hat{m} A)
\frac{1}{\partial_\tau + H}  .
\label{Eq:exp}
\end{equation}
Figure 2 shows the rotational $1/N_c$ corrections and $m_s$ ones 
diagrammatically.  
The rotational $1/N_c$ corrections are of particular importance
in the present model.  The time-ordering of collective operators 
has to be taken into account in the quantization,
since they do not not commute in principle.  The rotational $1/N_c$
corrections made it possible to solve the long-standing problem
of underestimation of axial charges and magnetic moments in the
$\chi$QSM~\cite{WW,Chretal1}.  In addition, there have been
arguments that the rotational $1/N_c$ corrections violate charge
conjugation~\cite{Schechter}.  However, Christov {\em et al.}
~\cite{ChrPobGo} and Wakamatsu~\cite{Wak} proved that the rotational
$1/N_c$ corrections possess correct properties under charge conjugation.
Moreover, it was shown by Prasza\l owicz {\em et al.}~\cite{PraBloGo2} 
that with the rotational $1/N_c$ corrections considered the $\chi$QSM
reproduces the result of the axial charge in the nonrelativistic
quark model, {\em i.e.} $5/3$, while it gives $1$ without them.

The collective SU(3) Hamiltonian is no longer SU(3) symmetric when
the $m_s$ corrections are taken into account.  Hence, the eigenstates
of the Hamiltonian are neither in a pure octet nor in a pure decuplet but
in mixed state with higher representations.
Dealing with the $m_s$ as a perturbation, we can obtain the mixed 
SU(3) baryon states:
\begin{equation}
|8,B\rangle \\;=\; | 8,B \rangle \;+ \; 
c^{B}_{\bar{10}} | \bar{10},B \rangle
\;+\;c^{B}_{27} | 27,B \rangle.
\label{Eq:wfc}
\end{equation}
The coefficients $c^{B}_{\bar{10}}$ and $c^{B}_{27}$ can be found 
elsewhere (see for example the recent review~\cite{Review}).

\section{Results and Discussion}

I want to remark the parameters of the model for the numerical
calculation, before I start to discuss the results.
The present SU(3) $\chi$QSM contains four free parameters.  
Two of them are fixed in 
the meson sector by adjusting them to the pion mass, 
$m_\pi=139\ \mbox{MeV}$, the pion decay constant, 
$f_\pi=93\ \mbox{MeV}$, and the kaon mass, $m_{\rm K}=496 \ \mbox{MeV}$.  
As for the fourth parameter, {\em i.e.} the constituent mass $M$
of up and down quarks, values around $M=420\ \mbox{MeV}$ have been used
because they have turned out to be the most appropriate one for the
description of nucleon mass splittings and other observables of baryons
(see ref.~\cite{Review}).  Hence, we fix the parameter $M$ to
420 MeV. For the description of the baryon sector, 
the method of Ref.\cite{Blotzetal} is chosen,
modified for a finite meson mass.   
The resulting strange current quark mass comes out around 
$m_s=180\ \mbox{MeV}$. 
With this set of fixed parameters, all the results which will be
presented from now on have been calculated.

\subsection{Electromagnetic Form Factors}
Setting $\hat\Gamma = \gamma_\mu$ and 
$\hat{O}=(\lambda_3+\lambda_8/\sqrt{3})/2$, we can calculate the 
electromagnetic form factors~\cite{Kimetal}.  
Figure 3 shows the electric form factors of the nucleon.
In the case of the proton, the result agrees well with the empirical data
~\cite{hoehler}.  
 As for the neutron, the result seems to be smaller than the empirical data
by Platchkov~\cite{pl}.  However, compared to the recent 
experiment conducted in Mainz~\cite{meyer}, the SU(3) result
is in good agreement with it.  The electric charge radii of the 
proton and the neutron are  $\langle r^2\rangle_{p}=0.78\;\mbox{fm}^2$  
and $\langle r^2 \rangle_{n} = -0.09 \;\mbox{fm}^2$, respectively. 
The corresponding experimental data are 
$\langle r^2\rangle_{p}=0.74\;\mbox{fm}^2$  
and $\langle r^2 \rangle_{n} = -0.11\pm0.003\;\mbox{fm}^2$~\cite{Kopecky}.  

In dotted curves in Fig. 1, the prediction of the SU(2) model is
shown.  As for the electric form factor of the proton, it is comparable
to the SU(2), whereas a great discrepancy is observed in the case of
that of the neutron.  
It is partly because of the absence of $m_s$ and 
terms appearing only in SU(3) and partly because of the different
expectation values of the collective operators. 

Figure 4 displays the magnetic form factors of
the nucleon.  
As we can see, the momentum dependence of the 
magnetic form factors are well reproduced, compared to the empirical data.
The $m_s$ corrections enhance the magnetic form factors about $10\%$ in
the case of the neutron, which is not negligible to improve the
prediction. 
 The magnetic moments of the proton and the neutron are, respectively,
$\mu_p=2.39\;\mu_N$ and $\mu_n=-1.76\; \mu_N$, while experimental data are
$\mu_p=2.79\;\mu_N$ and $\mu_n=-1.91\;\mu_N$.
The magnetic charge radii of the proton and the neutron are obtained as
follows: $\langle r^2 \rangle_p = 0.70$ $\mbox{fm}^2$ 
and $\langle r^2 \rangle_n = 0.78$ $\mbox{fm}^2$.
The corresponding experimental data are 
$\langle r^2 \rangle_p = 0.74$ $\mbox{fm}^2$ 
and $\langle r^2 \rangle_n = 0.77$ $\mbox{fm}^2$.
On the whole, They are in remarkable agreement with the experimental
data within around $15\%$.

In fact, one can show that any model
with {\em hedgehog} symmetry cannot reproduce the experimental
data of baryonic magnetic moments better than the error of $15\;\%$
\cite{KimPolBloGoe}.  In that sense, the $\chi$QSM lies in the upper 
limit of accuracy which can be attained in any hedgehog model in the
case of the magnetic properties.

\subsection{Scalar Form Factors}
It is of great interest to study the scalar form factor of the nucleon
\cite{KimBloSchGo},
since it provides also a clue of strangeness in the nucleon.
The analysis of the $\sigma_{\pi N}$ term the momentum dependence of
the scalar form factor was carried out by Gasser, Leutwyler and
Sainio~\cite{GaLeuSa}.  The results of Ref.~\cite{GaLeuSa} are
summarized as $\sigma = 45\pm 8$ MeV and $\Sigma\simeq 60$ MeV, and
$y= 2\langle N| \bar{s}s |N\rangle / \langle N| \bar{u}u+\bar{d}d |N\rangle
\simeq 0.2$ which means a share of $\langle N| \bar{s}s |N\rangle$
in the $\sigma_{\pi N}$ term.  The results indicate that
the strangeness content of the nucleon in the scalar channel 
is not negligible, while a couple of recent theoretical works
insist that there is no need to introduce a portion of strange quarks
to explain the $\sigma$ term~\cite{Bass,BallForteTigg}.  However,
though it might be small, it is still important to consider the
contribution of the strange quarks to the $\sigma$ term
in accordance with the recent experimental indication that
strange quarks might play an important role of explaining the properties of
the nucleon.  

Figure 5 draws the scalar form factor of the nucleon.
The error bar presented in Fig.5 stands for the empirical analysis
due to Gasser, Leutwyler and Sainio.  As shown in Fig. 5 our 
theoretical prediction is in good agreement with Ref.~\cite{GaLeuSa}.
The $m_s$ corrections seem to be negligible in the scalar form factor.
However, they play an important role of suppressing the strangeness 
contribution:
The value of $y$ with the $m_s$ corrections is
0.27 while $y=0.48$ without the $m_s$ corrections.  
It implies that though the $m_s$ corrections have a tiny effect on
the magnitude of the scalar form factor, it leads to a large suppression
of the $y$.The difference $\Delta \sigma = \sigma (2m_\pi^{2}) 
-\sigma (0)$ we have obtained is $18.18$ MeV.  This value is very 
close to what Ref.\cite{GaLeuSa} extracted, {\em i.e.} 
$\Delta \sigma = \sigma (2m_\pi^{2}) - \sigma (0) = 15.2\pm 0.4$ MeV.
The prediction of the $\chi$QSM for the scalar radius 
$\langle r^2\rangle^{S}_{N}$ is $1.5\; {\rm fm}^2$  which is also
comparable to the empirical value obtained by Ref.~\cite{GaLeuSa}
$\langle r^2\rangle^{S}_{N}\simeq 1.6 {\rm fm}^2$.

\subsection{Strange Vector Form Factors}
The strangeness content of the nucleon in the vector channel
is one of hot issues these days.  While a great deal of theoretical
works pile up, there is no clear theoretical consensus and no
evident experimental judgment yet.  

Encouraged by successful results of the scalar, electromagnetic and axial
properties in the SU(3) $\chi$QSM, it is of great interest to 
investigate the strange vector form factors in the same framework
\cite{KimWatGo}.
However, before we pursue the study of the strange vector form factors,
we need to take into account the kaonic effect properly in line
with the recent theoretical calculations incorporating 
the kaonic loops~\cite{Musolf}.  
 From this theoretical point of view, the strangeness 
in the nucleon can be interpreted in terms of the 
$K\Lambda$ or $K\Sigma$ components.  Figure 6 displays
a schematic diagram of explaining how the strangeness arises in the nucleon.
The diagram shown in the left-hand side of Fig. 6
can be redrawn in terms of quarks.
In fact, the nucleon is known to consist of three valence quarks,
{\em i.e.} $uud$.  When one of the $u$ quarks is hit by the 
external strange vector current as shown in Fig. 6,  
the $s\bar{s}$ pair is created and rearranged
so that we may have quark compositions $uds$ and $u\bar{s}$.    
They correspond respectively to $\Lambda$ and $K^+$.  
This explains that we have the contribution from the valence part
as well as the sea part to the strange vector form factors, 
though the nucleon itself does not include 
any strange valence quark.  

As is explained above, kaons play a dominant
role in describing the strange vector form factors.
The $\chi$QSM discussed up to now does not include the
kaon cloud which is believed to be of little importance in the
former calculation except for the neutron electric form factor
\cite{WatKimGo}.  However, to treat the strange vector form factors,
we have to take the effect of the kaon cloud into account properly.  
To do so, we have incorporated the kaonic tails 
selfconsistently in the profile
$P(r)$ in Eq.(\ref{Eq:profile}) at the expense of the pion tails. 

Figure 7 shows the strange vector form factors.  
The effect of the kaon cloud is in particular prominent in the case of 
the strange electric form factor.  As shown in
Fig. 7, the replacement of the pion cloud by the kaon one brings about
a sizable decrease of the strange electric form factor almost 
by a factor of three.  
 This remarkable result is in line with the recent investigation
of the kaonic effects on the neutron electric form factor~\cite{WatKimGo}.
 Such a drastic reduction of the strange electric form factor  
can be easily understood explicitly by evaluating the strange electric
radii.  In any hedgehog model the strange radii depend on the
inverse of the meson mass $\mu$ which suppresses the tail of the profile,
{\em i.e.} $\langle r^2\rangle_s \sim 1/\mu$.  From such a behavior
of the $\langle r^2\rangle_s$, we can derive the relation
\begin{equation}
\frac{\left. \langle r^2\rangle^{\rm Sachs}_{s} \right |_{\mu=m_\pi}}
{\left. \langle r^2\rangle^{\rm Sachs}_{s} \right |_{\mu=m_{\rm K}}}
= \frac{m_{\rm K}}{m_\pi} \simeq 3.5 .
\label{Eq:radius}
\end{equation}  
Eq.(\ref{Eq:radius}) explains the decrease of the 
$\langle r^2\rangle_{s}$ with $\mu=m_{\rm K}$.

Though the strange magnetic form factor is not changed as much as the
strange electric one, the effect of the kaon cloud is still 
noticeable.  In contrast to the strange electric form factor,
the kaon cloud enhances the magnitude of the 
strange magnetic one almost $50\%$.
As a result, by replacing the pion cloud by the kaon one,
the strange magnetic moment $\mu_s$ is brought from $-0.44$ $\mu_N$
to $-0.68$ $\mu_N$.
 
In addition, I want to mention a preliminary
experimental result of the strange magnetic moment 
announced by McKeown~\cite{Mckeown} in this workshop.
Surprisingly, his finding is {\em positive}.  It implies that  
it conflicts with almost all of the theoretical models 
including the present model.  
If the experimental result turns out to be correct, 
the present simple picture will not be enough to describe
the strange vector form factors.  
A more sophisticated and higher order corrections should be
considered.  However, we still anticipate more compiled
experimental data to enlighten us on it. 

\subsection{Tensor Charges}
Finally, I want shortly to present a recent calculation of the tensor
charges of the nucleon~\cite{su3ten}.  The tensor charge $\delta q$ is
related to the transverse quark distribution $h_1 (x)$.
However, $h_1(x)$ is not measurable in inclusive deep-inelastic
scattering.  That is the reason why it has not been extensively
studied for long.  In fact, Ralston and Soper~\cite{RalSop} proposed
the $h_1(x)$ which can be measured in polarized Drell-Yan processes
almost 20 years ago.  Recently, It was also suggested that
the $h_1(x)$ can be measured in other exclusive hard 
reactions~\cite{JaffeJi,Collins,BourSoffer}.

Jaffe and Ji demonstrated that the first moment of the $h_1(x)$ 
is related to the tensor charge of the nucleon:
\begin{equation}
\int^0_1 dx \left(h_1(x) - \bar{h}_1 (x)\right) = \delta q,
\end{equation}
where $\bar{h}_1 (x)$ is an antiquark transversity distribution.
Setting $\hat{\Gamma}=\sigma_{\mu\nu}$ and $\hat{O}=\lambda^a$
in Eq.(\ref{Eq:corr}), we can evaluate the tensor charges 
$g_{\rm T}^{a}$ in the $\chi$QSM, 
which are linearly related to $\delta q$.    
Note that the tensor charges depend on the renormalization
scale, though their dependence on it is
very weak.  The normalization point pertinent to the $\chi$QSM is
not determined from the first principle.  However, as mentioned in
section I, the renormalization point in the $\chi$QSM has be chosen
by $\rho^{-1} \simeq 600$ MeV, but there may be a factor of order unity.

In Tables I-II the results of the tensor charges are summarized.
As shown in Table I the rotational $1/N_c$ corrections are
of great significance numerically, while the $m_s$ corrections
are relatively small.  In contrast to the axial charges,
the tensor ones in the $\chi$QSM are closer to their values in the
nonrelativistic constituent quark model and in particular the 
strangeness contribution to the tensor charge $\delta s$ is compatible
with zero, while that to the axial ones $\Delta s$ in the $\chi$QSM
is negative and distinctive from zero~\cite{BloPraGo}.   
This difference between the tensor charges and the axial ones can
be explained by calculating them in the gradient expansion
and scrutinizing their dependence on the size of the soliton
\cite{Kimtensor}.

\section{Summary}
My aim in this talk has been to review the recent investigation
on various form factors of the nucleon 
in the SU(3) chiral quark-soliton model($\chi$QSM).
The rotational $1/N_c$ and linear $m_s$ corrections were taken into
account.  The only parameter we have in the model 
is the constituent quark mass $M$ which is fixed to 420 MeV by the
mass splitting of the SU(3) baryon octet and decuplet.
The results for the electromagnetic and scalar form 
factors are in remarkable agreement with experimental and empirical 
data within about 15 $\%$. 

The strange vector form factors were also discussed.  
The effect of kaon cloud turns out to be very large.  
By taking into account this kaonic effect, the strange charge radius 
is reduced to be almost 3 times smaller than that without the kaon cloud,
while the magnitude of the strange magnetic moment is increased.  

The tensor charges of the nucleon are presented in the same framework.
The $\chi$QSM predicts the number of the transversely 
polarized strange quarks $\delta s$ in the transversely polarized
nucleon compatible with zero, whereas it yields the negative nonzero 
number of the polarized strange quarks $\Delta s$ in the longitudinally 
polarized nucleon, which is consistent with the corresponding
experimental value.
 
\section*{Acknowledgments}
The works presented in this talk were done in collaboration with
A. Blotz, M.V. Polyakov, C. Schneider, T. Watabe and K. Goeke.
The work is supported by the BMBF, the DFG, and the COSY project 
(J\"ulich).  I am grateful to O. Benhar and A. Fabrocini for the
warm hospitality in Elba.
\begin{table}[]
\caption{Tensor charges $g^{(0)}_{\rm T}$, $g^{(3)}_{\rm T}$ and
$g^{(8)}_{\rm T}$ with the constituent quark mass $M=420\;\mbox{MeV}$.
The current quark mass $m_s$ is chosen as $m_s = 180$ MeV.
 The final model predictions are given by
${\cal O}(\Omega^1, m^1_{s})$.}
\begin{tabular}{cccc}
   & ${\cal O}(\Omega^0, m^0_{s})$ &  ${\cal O}(\Omega^1, m^0_{s}) $ &
${\cal O}(\Omega^1, m^1_{s})$ \\
\hline
$g^{(0)}_{\rm T}$     &   0  & 0.69 & 0.70 \\
$g^{(3)}_{\rm T}$     &   0.79  & 1.48 & 1.54 \\
$g^{(8)}_{\rm T}$     &   0.09  & 0.48 & 0.42 \\
\end{tabular}
\caption{Each flavor contribution to the tensor charges
as varying the constituent quark mass $M$.
The current quark mass $m_s$ is chosen as $m_s = 180$ MeV.}
\begin{tabular}{cccc}
$M$~[MeV]   & $\delta u$ &  $\delta d$  &  $\delta s$ \\
\hline
     420 &   1.12  & -0.42 & -0.008 \\
\end{tabular}
\end{table}

\vfill\break
\begin{center}
{\Large {\bf Figure Captions}}
\end{center}
\noindent
{\bf Figure 1}: The valence and sea contributions 
to an observable.  The left panel draws a schematic diagram for
the valence contribution, while the right one is for the sea contribution.
 The solid lines denote the $N_c$ valence quarks, whereas the loops
designate the polarized Dirac sea.  The wiggled line stands for
the corresponding external current $\Gamma$  to the observable 
$\langle \Gamma \rangle$.
\vspace{0.8cm}

\noindent
{\bf Figure 2}: The rotational $1/N_c$ and $m_s$ corrections.
The upper panel shows the rotational $1/N_c$ corrections, while
the lower panel is for the $m_s$ corrections.   
The solid lines denote the $N_c$ valence quarks, whereas the loops
designate the polarized Dirac sea.  The wiggled line stands for
the corresponding external current $\Gamma$  to the observable 
$\langle \Gamma \rangle$.
\vspace{0.8cm}

\noindent
{\bf Figure 3}: The electric form factors of the proton and
the neutron.  The left panel displays that of the proton, while the right
panel is for the neutron.  The solid curve corresponds to the strange quark
mass $m_s=180\;\mbox{MeV}$, while the dashed curve draws that without $m_s$.
The dotted curve shows the case of the SU(2) model.  $M=420\;\mbox{MeV}$
is chosen for the constituent quark mass.  The empirical data 
for the proton are taken from H\"ohler {\em et al.}~\cite{hoehler}
while for the neutron they are taken from Platchkov~\cite{pl}.
\vspace{0.8cm}

\noindent
{\bf Figure 4}: The magnetic form factors of the proton and
the neutron.  The left panel displays that of the proton, while the right
one is for the neutron.  The solid curve corresponds to the strange quark
mass $m_s=180\;\mbox{MeV}$, while the dashed curve draws that without $m_s$.
The dotted curve shows the case of the SU(2) model.  $M=420\;\mbox{MeV}$
is chosen for the constituent quark mass.  The empirical data are taken
from H\"ohler {\em et al.}~\cite{hoehler}.  The experimental
data for the neutron magnetic form factor (open triangle) are due to
Bruins {\em et al.}~\cite{Bruins}.
\vspace{0.8cm}

\noindent
{\bf Figure 5}: The scalar form factor of the nucleon 
$\sigma(t)$.  The solid curve displays the case of $m_s=148.49$ MeV, 
while the dashed curve shows that without $m_s$.  The error bar denotes
the empirical value from Gasser {\em et al.}~\cite{GaLeuSa}.
\vspace{0.8cm}

\noindent
{\bf Figure 6}: A schematic diagram of the kaon effect on the 
strange vector form factors.  The left panel displays the 
$\Lambda K^+$ component of the strange vector form factors,
while the right one shows the same component in a quark language.
\vspace{0.8cm}

\noindent
{\bf Figure 7}: The strange vector form factors of the 
nucleon.  The left one displays the strange electric form factor,
while the right panel draws the strange magnetic form factor.  
 The solid curve corresponds to the $\mu = m_{\rm K}$, while
dashed curve draws $\mu = m_\pi$.  
The constituent quark mass $M$ and $m_s$ are 420 MeV and 180 MeV,
respectively.

\vfill\break
\begin{center}
{\Large {\bf Figures}}
\end{center}
\vspace{1.6cm}
\centerline{\epsfysize=2.6in\epsffile{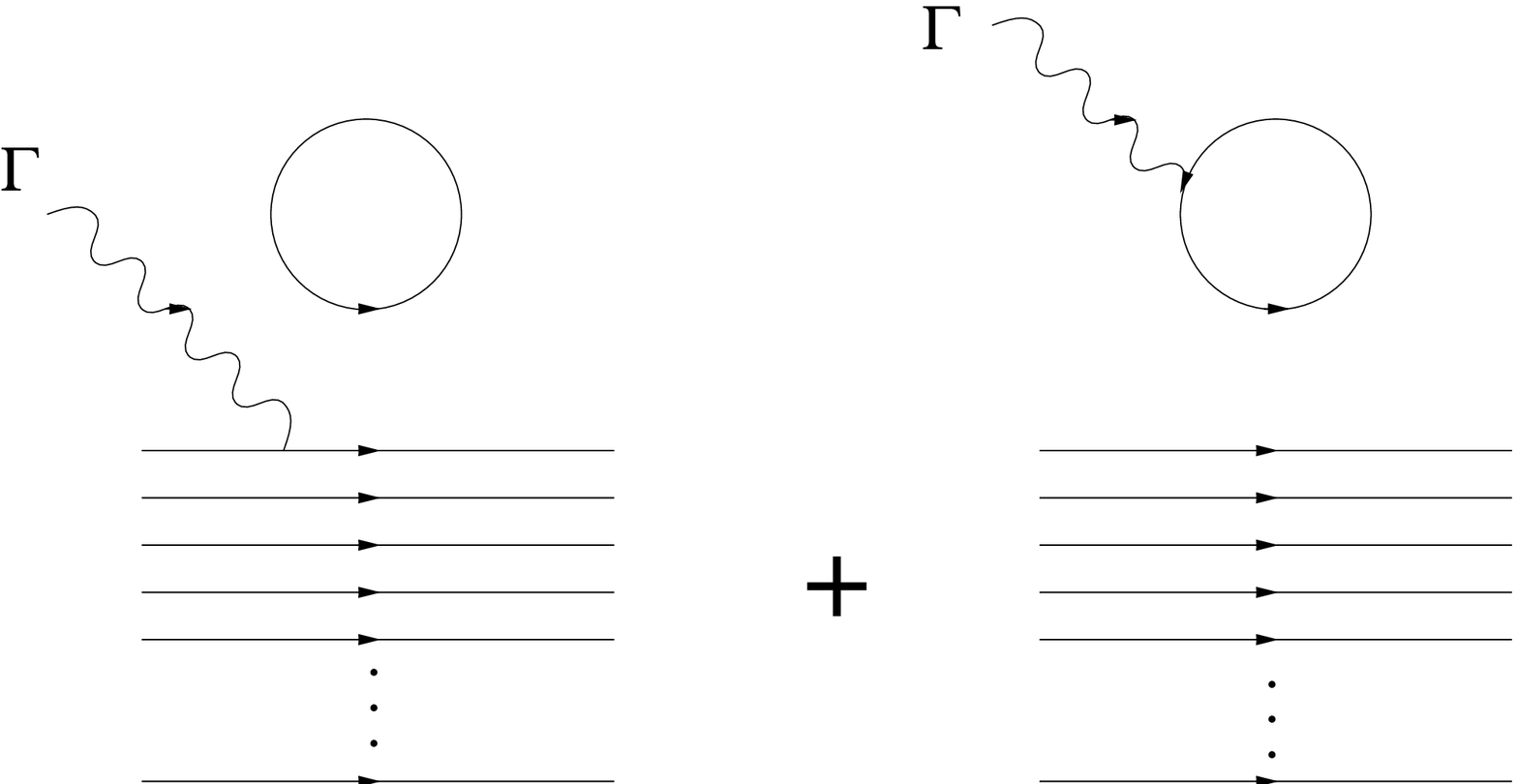}}\vskip4pt
\noindent \begin{center}	 {\bf Figure 1} 	 \end{center}   

\vspace{1.6cm}
\centerline{\epsfysize=3.7in\epsffile{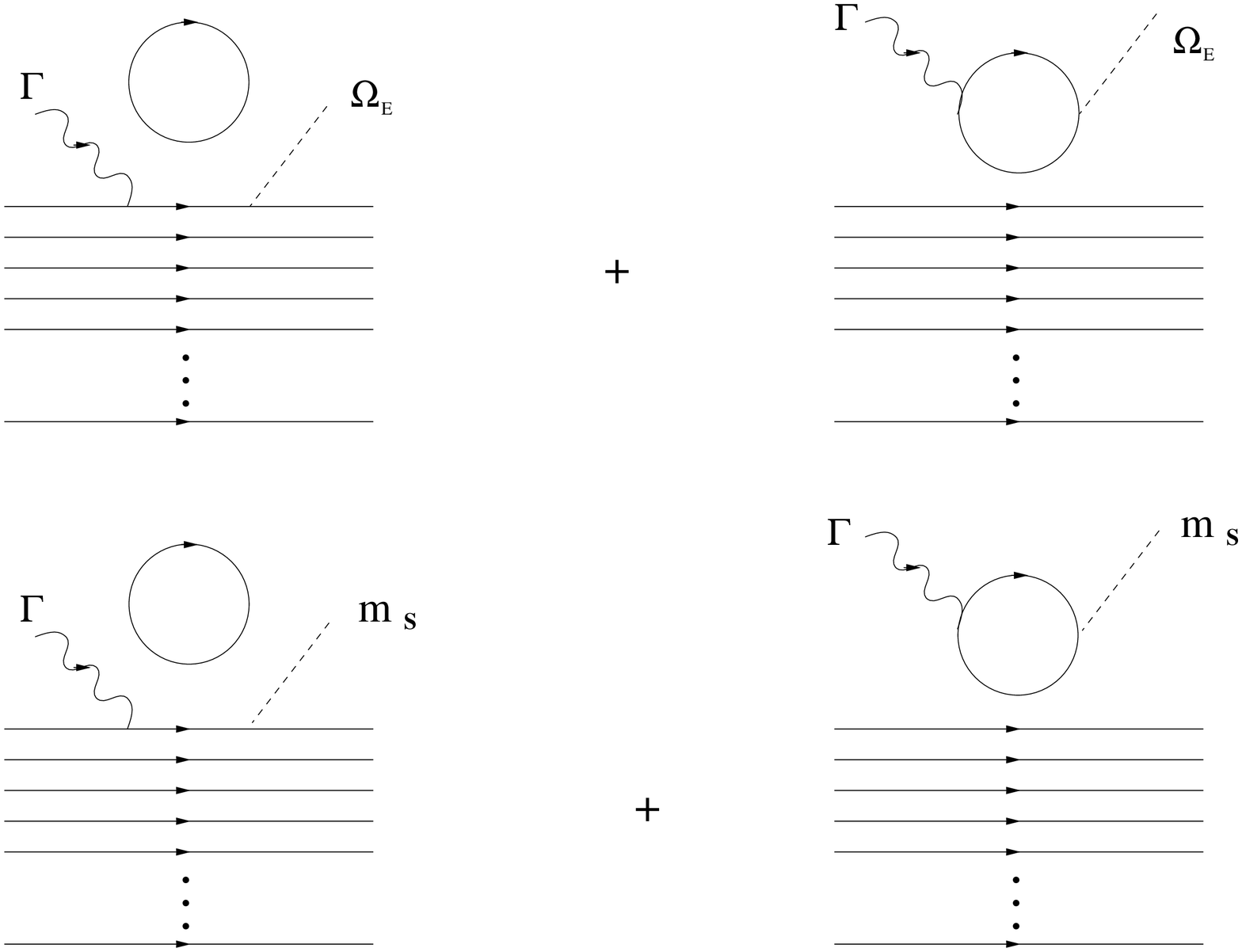}}\vskip4pt
\noindent \begin{center}	 {\bf Figure 2} 	 \end{center}   

\vspace{1.6cm}
\centerline{\epsfysize=2.7in\epsffile{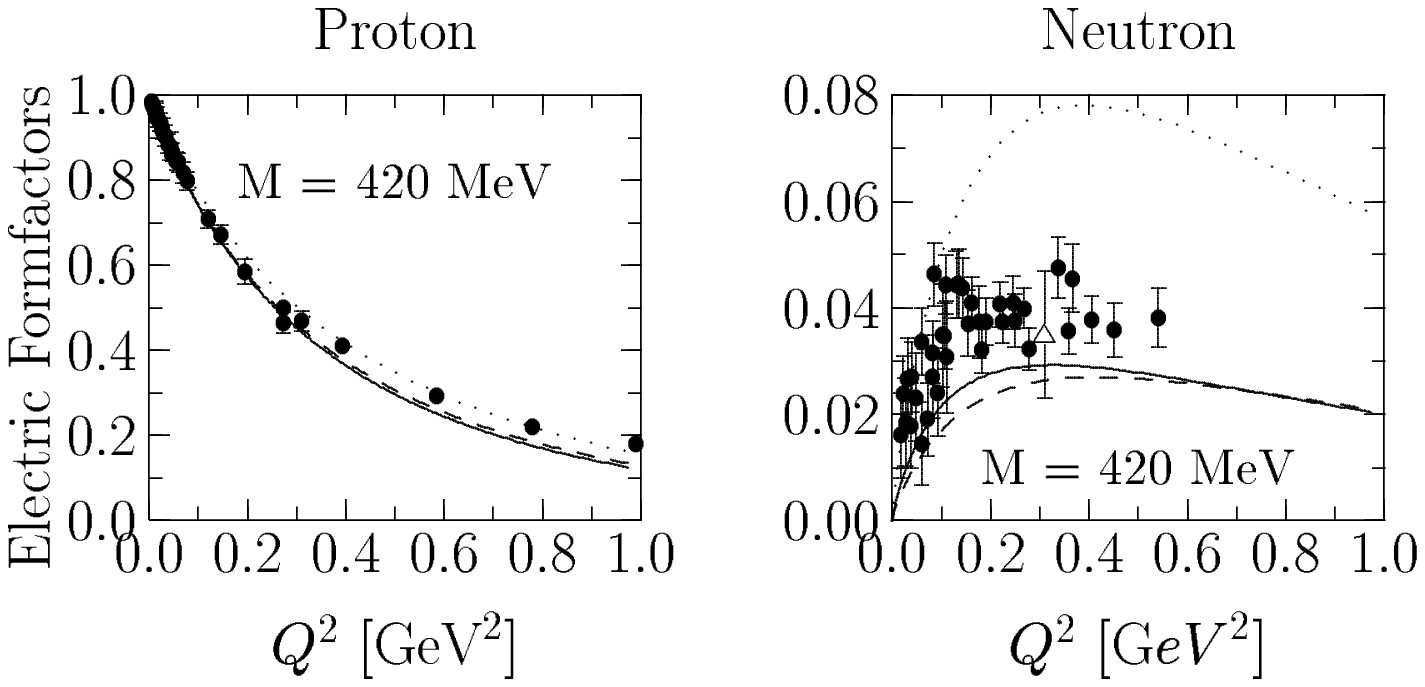}}\vskip4pt
\noindent \begin{center}	 {\bf Figure 3} 	 \end{center}   

\vspace{1.6cm}
\centerline{\epsfysize=2.7in\epsffile{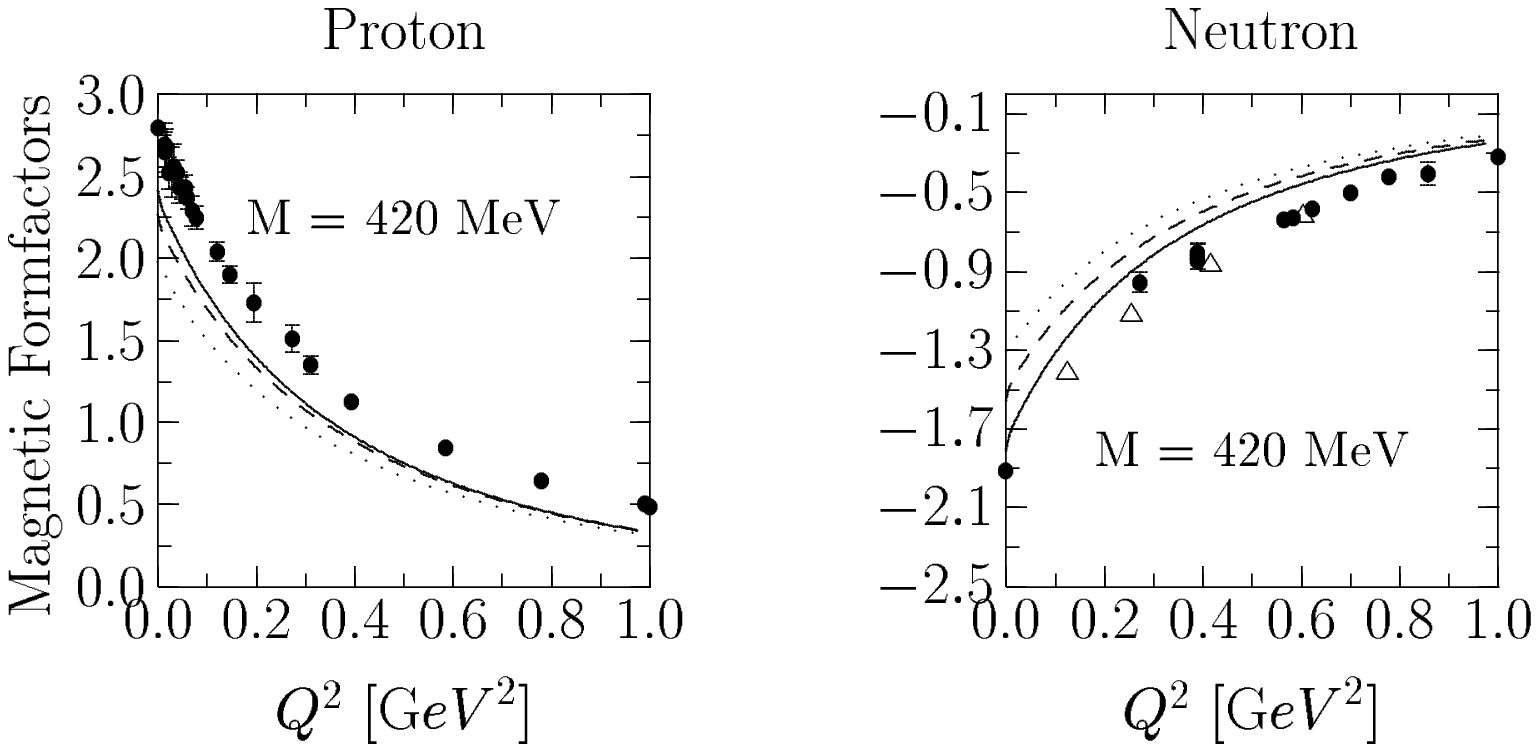}}\vskip4pt
\noindent \begin{center}	 {\bf Figure 4} 	 \end{center}   

\vspace{1.6cm}
\centerline{\epsfysize=3.2in\epsffile{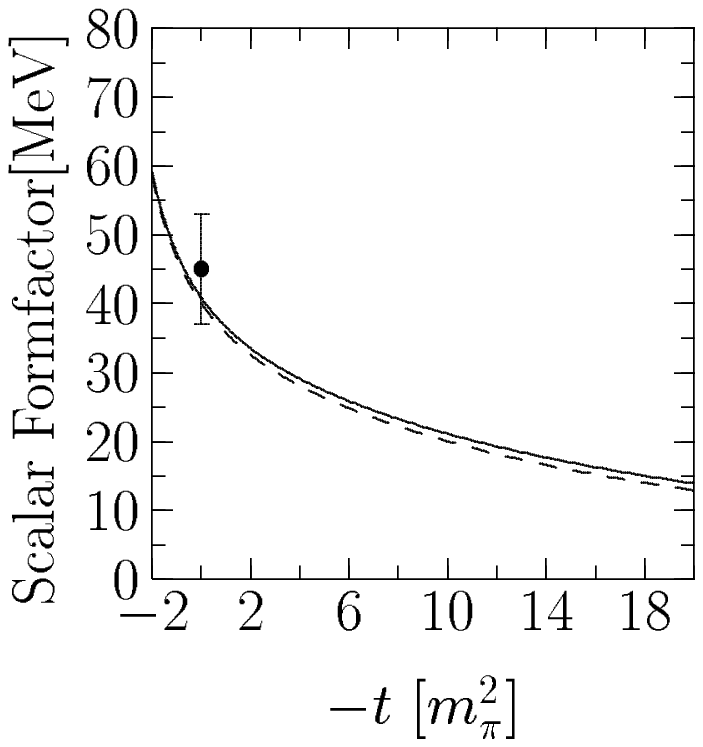}}\vskip4pt
\noindent \begin{center}	 {\bf Figure 5} 	 \end{center}   

\vspace{1.6cm}
\centerline{\epsfysize=2.4in\epsffile{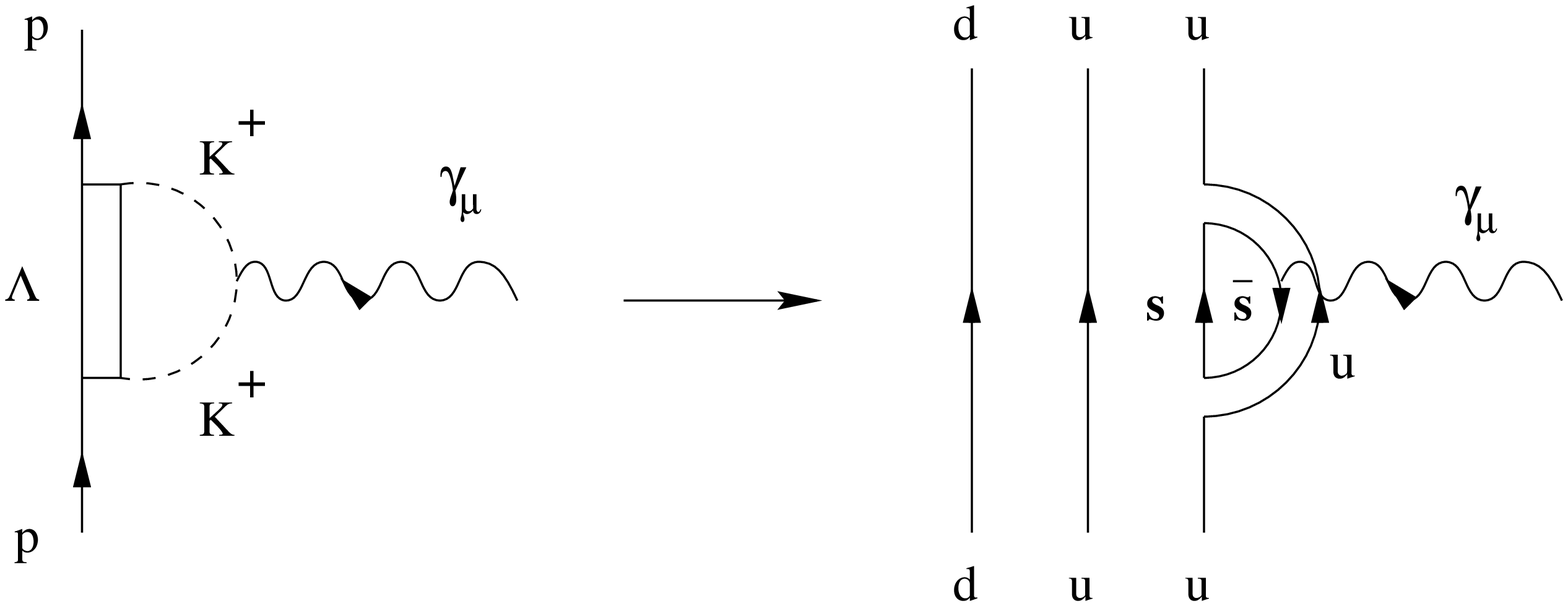}}\vskip4pt
\noindent \begin{center}	 {\bf Figure 6} 	 \end{center}   

\vspace{1.6cm}
\centerline{\epsfysize=2.7in\epsffile{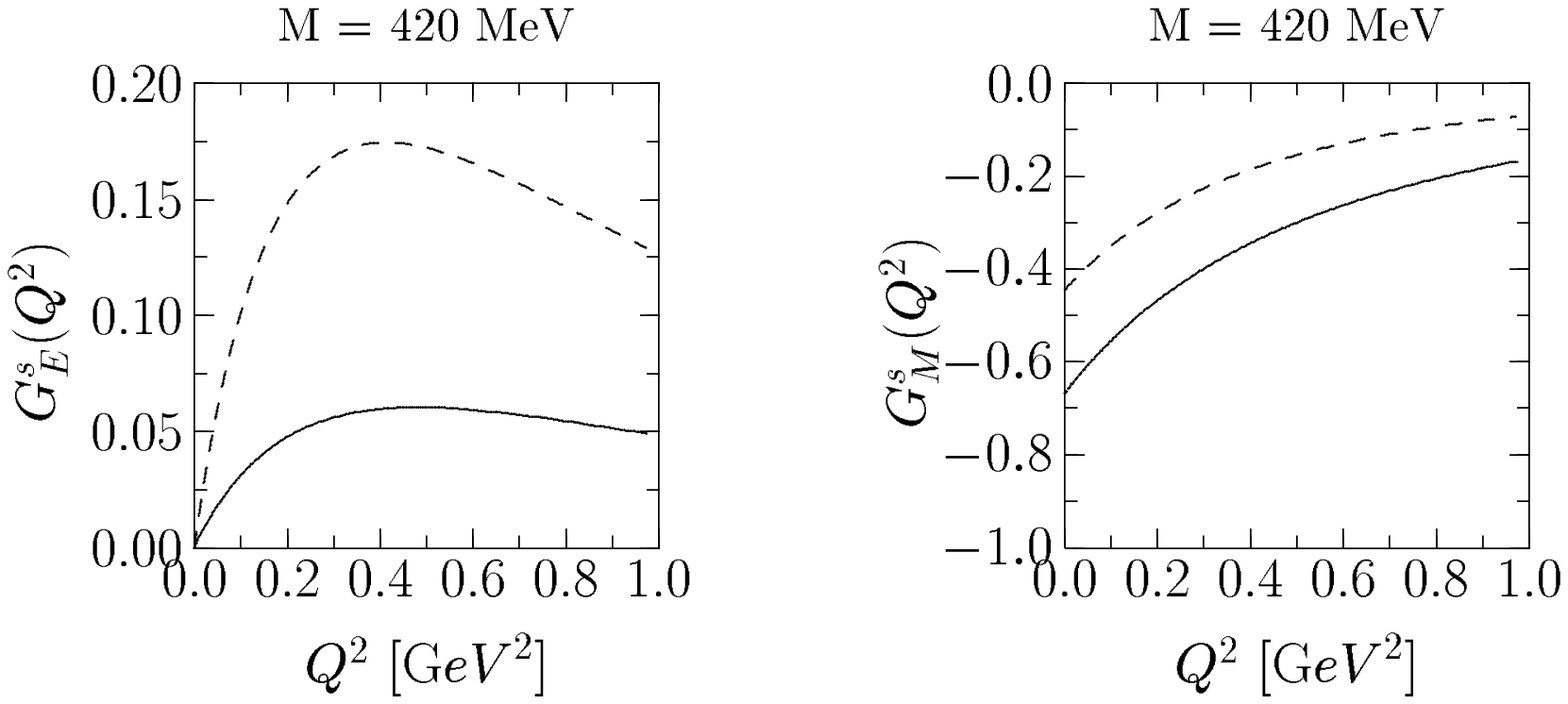}}\vskip4pt
\noindent \begin{center}	 {\bf Figure 7} 	 \end{center}

\end{document}